# Measuring Cognitive Abilities in the Wild:

# Validating a Population-scale Game-Based Cognitive Assessment

Mads Kock Pedersen[1,2], Carlos Mauricio Castaño Díaz[3], Qian Janice Wang[1,4], Mario Alejandro Alba-Marrugo[5], Ali Amidi[6], Rajiv Vaid Basaiawmoit[7], Carsten Bergenholtz[1], Morten H. Christiansen[8,9,10], Miroslav Gajdacz[1], Ralph Hertwig[11], Byurakn Ishkhanyan[12], Kim Klyver[13,14], Nicolai Ladegaard[15], Kim Mathiasen[15], Christine Parsons[10], Janet Rafner[1], Anders Ryom Villadsen[16], Mikkel Wallentin[9,10], Blanka Zana[1], and Jacob Friis Sherson[1,9,*]

1. Center for Hybrid Intelligence, Department of Management, Aarhus University, Aarhus, Denmark. 2. Department of Buisness Development and Technology, Aarhus University, Herning, Denmark. 3. Department of Architecture, Design and Media Technology, Aalborg University, Copenhagen, Denmark. 4. Department of Food Science, Aarhus University, Aarhus, Denmark. 5. Fundación universitaria Maria Cano, Medellín, Antioquia, Colombia. 6. Department of Psychology and Behavioural Sciences, Aarhus University, Aarhus, Denmark. 7. Faculty of Natural Sciences, Aarhus University, Aarhus, Denmark. 8. Department of Psychology, Cornell University, Ithaca, New York, United States of America. 9. School of Communication and Culture, Aarhus University, Aarhus Denmark. 10. Interacting Minds Centre, Aarhus University, Aarhus, Denmark. 11. Center for Adaptive Rationality, Max Planck Institute for Human Development, Berlin, Germany. 12. Department of Nordic Studies and Linguistics, University of Copenhagen. 13. Department of Entrepreneurship & Relationship Management, University of Southern Denmark, Kolding, Denmark. 14. Entrepreneurship, Commercialization and Innovation Centre (ECIC), University of Adelaide, Adelaide, Australia. 15. Department of Clinical Medicine – Department of Affective Disorders, Aarhus University Hospital, Aarhus, Denmark. 16. Department of Management, Aarhus University, Aarhus, Denmark. * Corresponding Author: sherson@mgmt.au.dk, Fuglesangs Allé 4, 8210 Aarhus V, Denmark, Cell Phone: +4528775765

## Abstract

Rapid individual cognitive phenotyping holds the potential to revolutionize domains as wide-ranging as personalized learning, employment practices, and precision psychiatry. Going beyond limitations imposed by traditional lab-based experiments, new efforts have been underway towards greater ecological validity and participant diversity to capture the full range of individual differences in cognitive abilities and behaviors across the general population. Building on this, we developed Skill Lab, a novel game-based tool that simultaneously assesses a broad suite of cognitive abilities while providing an engaging narrative. Skill Lab consists of six mini-games as well as 14 established cognitive ability tasks. Using a popular citizen science platform (N = 10725), we conducted a comprehensive validation in the wild of a game-based cognitive assessment suite. Based on the game and validation task data, we constructed reliable models to simultaneously predict eight cognitive abilities based on the users' in-game behavior. Follow-up validation tests revealed that the models can discriminate nuances contained within each separate cognitive ability as well as capture a shared main factor of generalized cognitive ability. Our game-based measures are five times faster to complete than the equivalent task-based measures and replicate previous findings on the decline of certain cognitive abilities with age in our large cross-sectional population sample (N = 6369). Taken together, our results demonstrate the feasibility of rapid in-the-wild systematic assessment of cognitive abilities as a promising first step towards population-scale benchmarking and individualized mental health diagnostics.

## 1. Introduction

Individual cognitive phenotyping holds the potential to revolutionize domains as wide-ranging as personalized learning, employment practices, and precision psychiatry. To get there, it will require us to rethink how we study and measure cognitive abilities. Much of what cognitive and behavioral scientists know about cognitive abilities and psychological behavior has been gleaned from studying small, homogeneous groups in the laboratory. Recent pushes to increase the number and diversity of participants (Bauer, 2020) are revolutionizing standards for power and generalizability across the cognitive and behavioral sciences. These advances have been enabled in part by moving from in-person testing to online equivalents, which are less costly for experimenters and more convenient for participants (Birnbaum, 2004). The maturation of these tools will be critical to realizing the promise of individual cognitive phenotyping, customizable diagnostics, and a revamp of intelligence research in general.

Going online with more convenient digital versions of traditional tasks makes it possible to scale up participant recruitment via crowdsourcing. Examples include projects such as LabintheWild (Reinecke & Gajos, 2015), Volunteer Science (Radford et al., 2016), and TestMyBrain (Germine et al., 2012), which offer a broad suite of digitized tasks from cognitive and behavioral science to volunteers from the general public. The success of these scientific platforms' in crowdsourcing data from customizable tasks has established them as a fruitful alternative to laboratory studies.

Online digital participation also allows for the possibility of developing novel forms of cognitive assessment that are gamified. Gamified assessment offers the potential to engage larger and more diverse participant pools in cognitive experiments than traditional tasks and, thus, amplifies the benefits of online crowdsourcing (Baniqued et al., 2013; Lumsden, Edwards, et al., 2016). Part of the allure of adding the gamified assessment to crowdsourcing is that it motivates players by framing the activity as an entertaining and playful way to contribute to a meaningful scientific question (Jennett et al., 2014; Sagarra et al., 2016).

The gamified approach can take different directions. In one direction, the traditional task for measuring cognitive abilities is preserved as much as possible, and game-like elements, such as graphics, points, and narratives, are added to frame the task as a game. Lumsden, Skinner, et al. (2016) is an excellent example of this, where the Go/No-Go task is gamified by adding wild west illustrations and framing the task as a game, where the villains should be shot and the innocent left alive. These game-like tasks have been shown to be more engaging, at least according to players' self-report, compared to their more traditional counterpart while producing similar results (Hawkins et al., 2013).

In another direction, new games are designed through an *evidence-centered design process*, whereby assessment tasks are designed to evoke behaviors that reveal targeted competencies (Mislevy et al., 2003). By designing a complete game from scratch around specific cognitive abilities, researchers can obtain richer information than the traditional pen and paper version (Hagler et al., 2014). The games can be more complex and dynamic, which allows for more interesting cognitive modeling (Leduc-McNiven et al., 2018). Moreover, cognitive assessment games often apply *stealth assessment* (Shute et al., 2016), where the cognitive ability measures are derived from the players' in-game behavior. Thus, the players are immersed in the game experience rather than being constantly aware of being tested (Shute et al., 2016; Valladares-Rodríguez et al., 2016).

Prominent examples of games built for cognitive assessment and applied at a large scale are *Sea Hero Quest* (Coughlan et al., 2019) and *The Great Brain Experiment* (H. R. Brown et al., 2014). Sea Hero Quest delivers a casual game experience and has reached 2.5 million participants, which yielded important insights into spatial navigation impairments in adults at risk of Alzheimer's disease (Coutrot et al., 2018). That said, Sea Hero Quest is by design only intended to measure spatial navigation; thus, if the goal is to measure a portfolio of distinct cognitive abilities, it would be a considerable effort to perform similar studies for each cognitive ability of interest. In contrast, The Great Brain Experiment is a collection of smaller games that assess multiple cognitive abilities. Through a large-scale deployment, the games have yielded new insights into age-related changes in working memory performance (McNab et al., 2015) and patterns of bias in information-seeking behavior (Hunt et al., 2016). While demonstrating the viability of large-scale cognitive ability testing (H. R. Brown et al., 2014), the two above-mentioned studies have relied subsequently on small, laboratory-based samples to validate their gamified cognitive ability measures originally derived from large-scale data collection. Ideally, it would be preferred to have the same person play the game as well as perform the validation tasks. This, thus, raises an important question: How can robust within-subject validation of game-based cognitive ability measures be achieved by motivating large groups of players to both play the games as well as perform the less entertaining and more time-consuming traditional cognitive tasks?

Here, we present Skill Lab, an original suite of games that takes advantage of the demonstrated power of online recruitment to validate novel gamified assessments of a broad portfolio of cognitive abilities. Our comprehensive mapping of multiple abilities within the same game allows us to assess their interrelations, as well as correlations with participant demographic factors, in a broad cross-section of a national population. Finally, whereas this study is based on current theoretical considerations, the benefits of the gamified approach discussed above could, in the long run – when combined with appropriate clinical tests – provide the level of systematic mapping of cognitive and psychological demographics (e.g., central executive functioning or personality traits) and individualized profiling required towards population-scale benchmarking and individualized mental health diagnostics.

## 2. Game development

### 2.1 Theoretical considerations

With the aim to contribute new knowledge to the assessment of cognitive abilities in the wild, we designed an ambitious suite of games that would simultaneously test a broad set of cognitive abilities. This process started by identifying how cognitive abilities have been operationalized and measured in laboratories. From this literature search, we selected 13 cognitive abilities (Fig 1) suitable for gamification while ensuring broad coverage of important areas for everyday cognitive functioning (Lezak et al., 2012). To determine the suitability for gamification of a cognitive ability, we had several iterative workshop sessions with game designers in which we brainstormed game-mechanics that could activate the specific ability. The cognitive abilities we selected have generally been investigated as relatively distinct aspects of cognition: executive functioning, language, and visual function, with indications of more nuanced subcomponents (Carroll, 1993; Deary, 2011; Jensen, 1998; Knopik et al.,

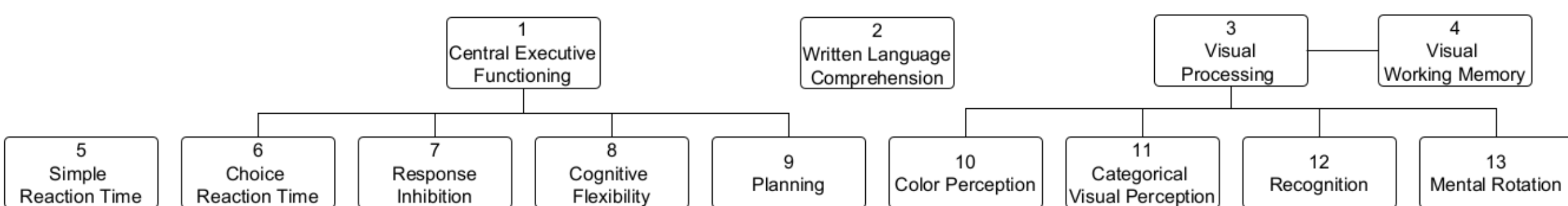

*Fig 1 The 13 cognitive abilities that we aim to measure through Skill Lab. The relationship between the cognitive abilities shown here are for illustrative purposes only. This is not a complete representation of all possible cognitive abilities, and we have not mapped all the possible relations between the components.*

2017; Mackintosh, 1998). Table 1 contains our descriptions for each of the 13 cognitive abilities (see Supplemental Information for overviews of all the tasks used to measure the different cognitive abilities – Section 3 and 4 – and how they are operationalized – Section 5 and 6).

*Table 1. Descriptions of each of the cognitive abilities. The relationship between the different abilities can be found in Fig 1.*

| Fig 1 Index | Cognitive Ability | Description |
|---|---|---|
| 1 | Central Executive Functioning | Central executive functioning has several definitions. It is proposed to include various cognitive functions, such as planning, inhibiting responses, developing strategies, flexible action sequencing, maintaining behavior. Essentially, central executive functioning consists of various classes of behavior used in self-regulation (Logan, 1985). Therefore, an executive act is any action toward oneself (whether conscious or not) that functions to change one's behavior to change future outcomes (Barkley, 2001). |
| 2 | Written Language Comprehension | Written language comprehension is the ability to process textual information. At the sentence level, processing involves many subcomponents, such as recognizing individual written words, understanding how the words relate to each other, how the words fit together in sentences, and how the context constrains the interpretation of the sentence (Rodd et al., 2015). |
| 3 | Visual Processing | Visual processing is the ability to perceive, process, analyze, and manipulate visual information and involves the storage and recall of visual representations via visual imagery and memory (Castro-Alonso & Atit, 2019). |
| 4 | Visual Working Memory | Visual working memory involves storing and maintaining visual information in the short term (L. A. Brown et al., 2006). |
| 5 | Simple Reaction Time | Simple reaction time refers to the time needed to respond to a single stimulus as quickly as possible. Performance on simple reaction time tasks very often correlates with the performance of other psychometric tests. It is believed to indicate cognitive processing speed and is one of the most basic measurements of cognitive performance, underlying all cognitive functions. Studies of reaction times are critical in studies about aging, as reaction times increase with age (Deary et al., 2011). |
| 6 | Choice Reaction Time | Choice reaction time involves making appropriate responses as quickly as possible when challenged with two or more response options. Choice reaction time captures aspects of processing speed under complex task conditions and shows a moderate to strong correlation with general fluid intelligence (Deary et al., 2011). |
| 7 | Response Inhibition | Response inhibition is the ability to stop oneself from performing an action when the action is no longer required or is inappropriate. Inhibiting one's responses is a component of executive |
| | | functioning, as it supports flexible and goal-oriented behavior in changing contexts (Verbruggen & Logan, 2008). |
| 8 | Cognitive Flexibility | Cognitive flexibility refers to shifting between different tasks depending on contextual demands and is a component of executive function. Cognitive flexibility is vital in life, as we are faced with situations that require multitasking or rapid task switching every day. When talking about cognitive flexibility, the concept of switching costs is fundamental. Studies show that switching back and forth between tasks can harm productivity. On the other hand, task switching can be beneficial when stuck, as it can increase creativity by decreasing cognitive fixation (Geurts et al., 2009; Lu et al., 2017; Monsell, 2003). |
| 9 | Planning | Planning refers to the ability to anticipate and plan actions, as well as to monitor goal attainment, and when necessary, update plans mid-execution. It involves a supervising function that is linked with the frontal activation and is essential for successful problem-solving (Dockery et al., 2009). Planning is often viewed as a subcomponent of executive functioning (Carlson et al., 2004; Krikorian et al., 1994). |
| 10 | Color Perception | Color perception is the ability to detect differences in stimuli with varying distributions of spectral energy. These differences must be based on the color's hue or saturation rather than the intensity contrast of the stimuli (Jacobs, 1993). |
| 11 | Categorical Visual Perception | Categorical visual perception refers to the ability to organize concepts (e.g., objects or attributes of objects) into distinct categories, with the consequence that cross-category stimuli will be more easily distinguishable than within-category stimuli (Harnad, 1987). |
| 12 | Recognition | Recognition is the ability to identify information - in this case about objects - from previous encounters or knowledge, such as shape or color. This is a cue-based, associative process and is related to visual search processes (Ullman, 2000). |
| 13 | Mental Rotation | Mental rotation is the ability to mentally rotate objects and scenes and recognize them when looking at them from various orientations. This skill is closely related to navigational skills (Collins & Kimura, 1997) and serveral discrete processes involved during visual search, transformation, and recognition (Xue et al., 2017). |

## 2.2 The game - Skill Lab

With the theoretical model as a starting point, we held multiple brainstorming sessions with game designers to identify game mechanics that could activate the different cognitive abilities. The game mechanics that were found during the brainstorming sessions were combined into six games through an evidence-centered design process: Rat Catch, Relic Hunt, Electron Rush, Shadow Match, Robot Reboot, and Chemical Chaos (Fig 2a–f, see Supplemental Information Section 7 for complete descriptions of the designs). These six games were collected into a single application called Skill Lab. Skill Lab contained an overarching structure and a detective narrative theme intended to motivate and guide the participant between the games. For this paper, we limit the scope of our analysis to the measures derived from participants' behavior within the six games and the validation tasks.

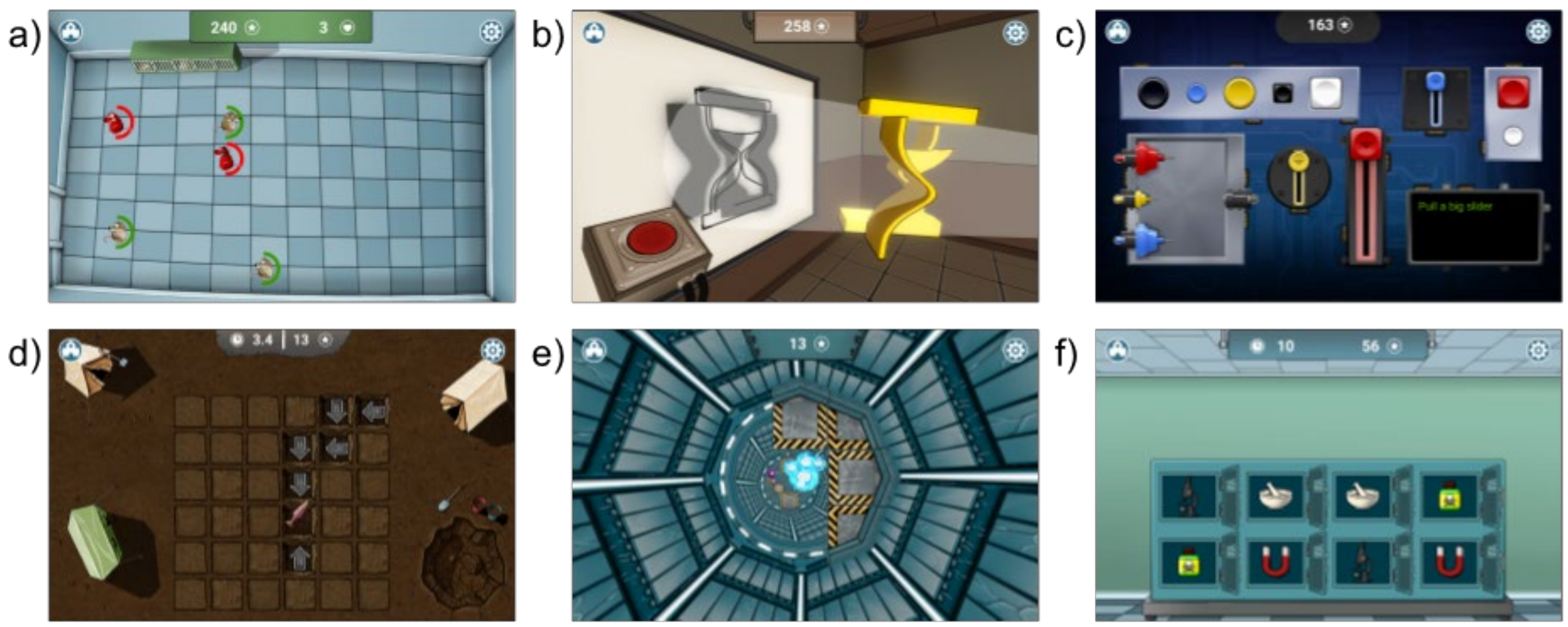

*Fig 2 The six games making up Skill Lab. a) Rat Catch is designed to test response inhibition, simple reaction rime, and choice reaction time, b) Shadow Match to test visuospatial reasoning in 3D, c) Robot Reboot to test reading comprehension and instruction following, d) Relic Hunt to test visuospatial reasoning and executive functions for simple strategy making in 2D visuospatial scenarios, e) Electron Rush to test how people navigate and make decisions, and f) Chemical Chaos to measure visual working memory.*

The games were designed to measure the cognitive abilities via stealth assessment (Shute et al., 2016). We created the games with the distinctive feel of a casual game while activating the targeted cognitive abilities. A consequence of this design choice is that the games are not a one-to-one redesign of any particular standard cognitive task. However, there are significant shared elements allowing connections to be drawn between the cognitive abilities most likely to be activated. We could, as an example, take the relationship between the classic Go/No-Go task (Lee et al., 2009) and the Rat Catch game (Fig 2b). The Go/No-Go task, typically administered in test batteries, measures response inhibition, simple reaction time, and choice reaction time (when facing distractors) by presenting a participant with a series of stimuli. If the stimulus is the correct type, the participant must react as quickly as possible; otherwise, the participant should refrain from reacting. This test procedure has an analog in the first two levels of Rat Catch. In the first level, a rat appears for a limited time at a random position; the player is asked to tap the rat as quickly as possible, providing simple reaction time measures. The rats disappear faster and faster as the level progresses. Once the player misses three rats, this level of play ends. In the second level of the game, there is a 50% chance that an "angry" red rat will appear. The player is instructed not to react to red rats but to still tap all other rats as quickly as possible. The level then follows the same progression as the first level, ending after three errors have been made (either tapping a red rat or not tapping the other rats before the timer runs out). This taps into choice reaction time and response inhibition. Further, Rat Catch levels add variations, such as an increasing number of stimuli or moving targets that have no analog in the Go/No-Go task. These additions give indicators of visuospatial reasoning components, such as 2D spatial representation and movement perception. Finally, relevant game indicators, such as average reaction time and accuracy in different levels, were identified via cognitive task analysis (Newell, 1966; Newell & Simon, 1972; Shute et al., 2016), where we mapped cognitive abilities required to achieve specific player behavior in the games (see Supplemental Information Section 8).

## 3. Methods

### 3.1 Participants

Participant engagement typically has an exponential fall off (Lieberoth et al., 2014), and in this case, a substantial player effort was needed to play both the games and complete the validation tasks; thus, broad and efficient recruitment was essential. Skill Lab was therefore launched publicly in Denmark in collaboration with the Public Danish Broadcast Company (Danmarks Radio, DR) on the 4th of September 2018 on (https://www.scienceathome.org/games/skill-lab-science-detective/, Retrieved: 2020-07-07), Apple Appstore, and Google Play. The Committee of Research Ethics for Region Midtjylland (Denmark) exempted the study from ethical oversight, and the project received ethical approval from the Institutional Review Board at Cornell University (Protocol ID: 1808008201). The study was conducted in accordance with all ethical requirements. Thus, the players provided informed consent before taking part in the study and any data were recorded. The players were made aware that they could, at any time, leave the study and request their data to be anonymized.

To attract the broadest possible audience, we drew attention to the project through a series of DR news articles with themes varying from AI and technology to psychology and computer games (https://www.dr.dk/nyheder/viden/nysgerrig/tema/danmarks-nye-superhjerne, Retrieved: 2020-07-07). Furthermore, Skill Lab was part of an educational event across classes at 208 high schools during the first week of December 2018. This event accounts for the spike of users at age 16 (Fig 3).

All in all, more than 16,000 people signed up to play the publicly available version. The game was available in versions running either on mobile devices or in the browser of personal computers. Since the required user interactions were different between mobile and computer versions, each version was separately validated (Drucker et al., 2013; Muender et al., 2019; Watson et al., 2013). This paper focuses primarily on the mobile version since it had the broadest reach. The results presented in the paper are based on the sample of 6,524 players from the in-the-wild data set that played at least one game on the mobile version. We also test the generated models on a sample of 4201 players from the in-the-wild data set that played at least one game on the computer version of which 603 also had cognitive abilities from the tasks.

The participants who played at least one game on the mobile version represent a broad cross-section of the Danish population (Danmarks Statistik, 2020) in terms of gender (3181 female, 3296 male, and 47 other[1]; or 49%, 50%, and 1%, respectively) and age (Fig 3), starting at age 16 years — the minimum age for granting informed consent according to the EU's General Data Protection Regulations. For demographic distribution of the computer players, the players in the validation sample, and the players not in the validations sample see Supplementary Information Section 9.

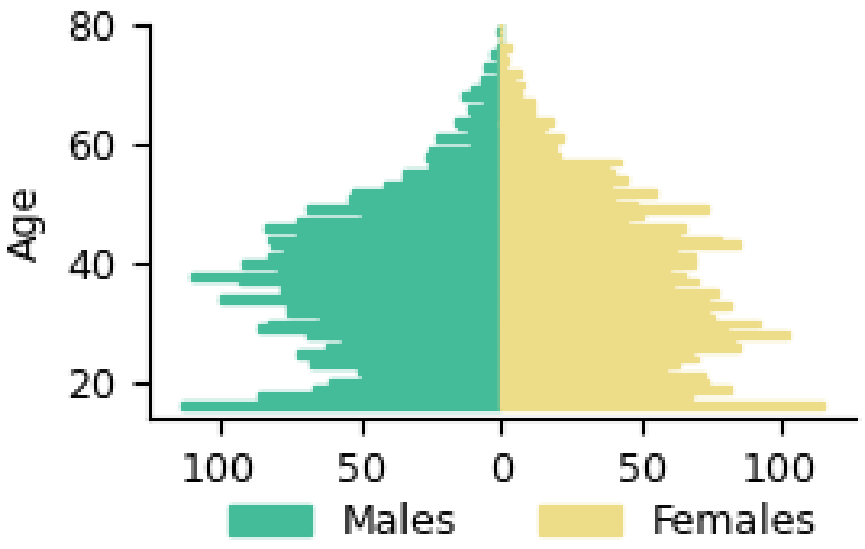

*Fig 3 Age distribution by gender for players who played at least one game in the wild. There are no qualitative differences in the age and gender distribution between those that played the game on mobile devices and those that played it on computers.*

### 3.2 Measuring convergent validity of the game-based cognitive measures

Many traditional cognitive tasks aim to assess a limited number of targeted cognitive abilities under strict conditions that minimize distractions and maximize experimental control (Salthouse, 2011). In contrast, the Skill Lab games are designed to engage multiple cognitive processes, simultaneously measuring multiple abilities within a convenient, engaging, and scalable package that aims to increase the external validity of the cognitive measures by creating a more realistic context and gameplay compared to traditional tasks (Schmuckler, 2001; Valladares-Rodríguez et al., 2016).

To test the convergent validity of the cognitive abilities' measures from the six games, we administered 14 standard cognitive tasks in a separate section of Skill Lab (see Supplemental Information Section 4 for full descriptions):

- Corsi Block (Kessels et al., 2000)
- Deary-Liewald (Deary et al., 2011)
- Eriksen-Flanker (Davelaar & Stevens, 2009)
- Groton Maze (Papp et al., 2011)
- Mental Rotation (Ganis & Kievit, 2015)
- Go/No-Go (Lee et al., 2009)
- Stop Signal (Verbruggen & Logan, 2008)
- Stroop (Zysset et al., 2001)
- Token Test (Turkyılmaz & Belgin, 2012)
- Tower of London (Kaller et al., 2011)
- Trail Making (Fellows et al., 2017)
- Visual Pattern (L. A. Brown et al., 2006)
- Visual Search Letters (Treisman, 1977)
- Visual Search Shapes (Treisman, 1977).

To obtain quantifiable measures of the players' ability levels, we identified *indicators* of the cognitive abilities assessed (e.g., number of errors in a task) in both the games (45 indicators, see Supplemental Information Section 8) and the tasks (68 indicators, see Supplemental Information Section 6). The game indicators were identified through a cognitive task analysis (Newell, 1966; Newell & Simon, 1972), whereby the stimuli in the games were connected to the corresponding actions a player could make and how the player's cognitive abilities could influence these actions. The full theoretical mapping between cognitive abilities, games, and validation tasks can be found in Fig 4.

Since many tasks conceptually measure aspects of the same cognitive abilities, combining the observations from different tasks with a strong theoretical overlap can give rise to more robust composite measures of cognitive abilities. Measures of cognitive abilities from tasks can be defined on a spectrum of computational granularity; pure indicators (Salthouse, 2011), linear combinations of indicators (Bollen & Bauldry, 2011), all the way to methods like generative models (Guest & Martin, 2020). Here, we form linear combinations of indicators, combining indicators from multiple tasks according to a standard theoretical interpretation, as it is the simplest way to take advantage of the overlap among the indicators. We recognize that the association between any particular combination of indicators is open to debate and offer the specific aggregation of indicators here as the most straightforward theoretical proposal. (For a list of the standard tasks indicators associated with each of the 13 cognitive abilities, see Supplemental Information Section 6).

[1] For the sake of transparency, are we are using the term ”other” here as that was the multiple-choice answer option given to the participant, rather than add an interpretation such as non-binary, non-conforming or any other term that would have been more appropriate.

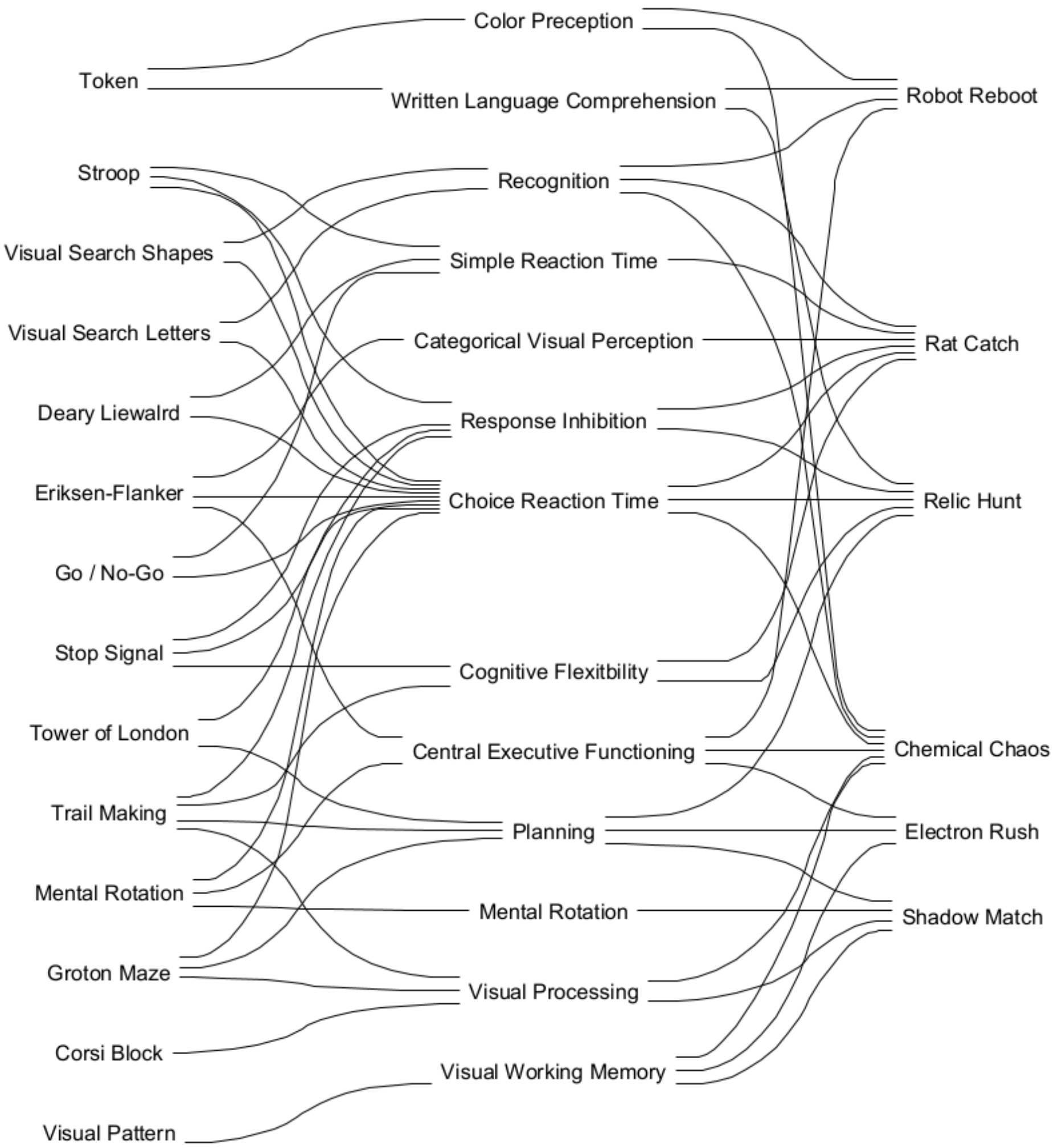

*Fig 4 Map of task, cognitive abilities, and game connection from a theoretical point of view. In the first column are all the tasks, in the second are all the cognitive abilities, and in the third are all the games. Each task measures a series of indicators informing about a cognitive ability. Each connection between the first and the second columns means that there is at least one indicator of a task informing about a cognitive ability. The connections between the second and the third column identify a theoretical link from the task analysis between a cognitive ability and a game.*

## 3.3 Modelling cognitive abilities with games and validation tasks

To be included in the validation process, a player had to complete at least one specific combination of validation tasks for a given cognitive ability. From 6369 players on the mobile version, we obtained a large sample of wild players (N=1,385) that had taken the right combination of validation tasks to measure at least one cognitive ability (e.g., the three tasks Visual Pattern, Groton Maze, and Corsi Block had to be completed for us to evaluate the ability visual working memory).

We trained a linear model that uses game data to predict players' cognitive abilities, where cognitive abilities are operationalized by measurements from the validation tasks (Fig 5). We started by defining cognitive ability measures by combining indicators - that measure the same construct - from different tasks. To determine which indicators to combine, we reviewed the tasks and identified the indicators $t_i$ (see Theoretical Considerations) of a cognitive ability that had a theoretical overlap (Beaujean & Benson, 2019; Mayo, 2018). For each of the 68 task indicators $t_i$, we assigned 13 coefficients $\alpha_{ij} \in \{-1,0,1\}$ depending on its theoretical contribution to each of the cognitive abilities $C_j$ by assigning: 0 if there is no contribution, 1 if there is a positive correlation between the task indicator and the cognitive ability, and -1 if there is a negative correlation (see Supplemental Information Section 5 for a comprehensive list of coefficients and justifications). The task indicators were standardized and combined into measures of cognitive abilities (Bollen & Bauldry, 2011) by taking weighted averages

$$C_j = \frac{\sum_{i=1}^{68} \alpha_{ij} t_i}{\sum_{i=1}^{68} |\alpha_{ij}|}.$$

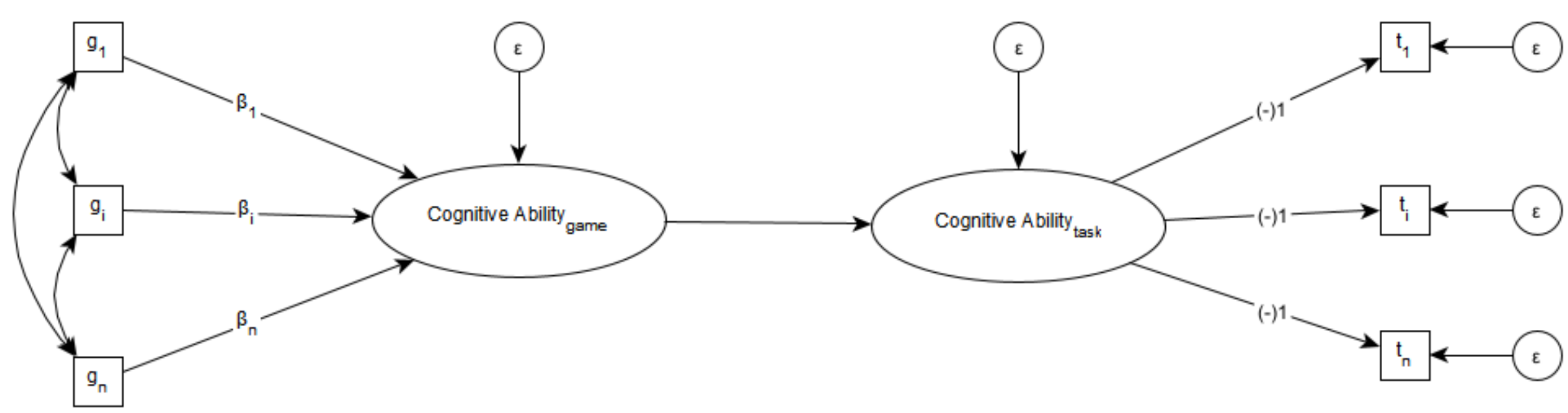

*Fig 5 Illustration of the predictive model that we test. On the right-hand side, the cognitive abilities from tasks are theoretical constructed via reflective indicators from the tasks with weights (-1 or 1) derived from theory. On the left-hand side cognitive abilities from games are estimated using the elastic net that handles the collinearities between the game indicators.*

For the games, we identified 45 indicators $g_i$ from the six games that contained information pertaining to the cognitive abilities. Before any modeling was performed, all game indicators and cognitive ability measures were standardized to mean = 0 and SD = 1. Only players who had produced all the task indicators associated with the respective cognitive ability (see Supplemental Information Section 6) and at least one game indicator were included in the sample used to fit the linear regression models predicting the cognitive abilities measured from the tasks with game indicators (for sample sizes see Table 3). Any missing game indicators were imputed using multivariate imputation with chained equations (Buuren & Groothuis-Oudshoorn, 2011), which generated one common imputation model for the entire data set. The imputation model was generated from game indicators only and contained no information about task indicators or demographic information. To prevent overfitting, an elastic-net model (Zou & Hastie, 2005) was used.

Elastic-net models combines ridge (Hoerl & Kennard, 1988) and lasso (Tibshirani, 1996) regression by adding two penalty terms (regularization) to the loss function when fitting the coefficients of a linear model

$$\hat{\beta} \equiv \underset{\beta}{\operatorname{argmin}}(\|y - X\beta\|^2 + \lambda_2\|\beta\|^2 + \lambda_1\|\beta\|_1),$$

where $\beta$ are the coefficients of a linear model, and $\lambda_1, \lambda_2$ are determining how much of the, respectively, lasso and ridge penalties to apply. Both ridge and lasso regression prevent overfitting; ridge regression by shrinking the values of the colinear coefficients closer to zero, i.e., grouping colinar game indicators, and lasso by forcing some of the coefficients to be exactly zero, i.e., automatic variable selection. The elastic net model increases the reliability of the model over ordinary least squares regressions, as it can handle multi-collinearity among the indicators by shrinking the coefficients or zeroing redundant indicators. Thus, one must be careful when interpreting the coefficients resulting from the elastic-net model as a small or zero coefficient could be either a redundant or irrelevant indicator, and therefore, not an unequivocally sign that the indicator contains no information about the cognitive ability. As our focus is to generate a predictive model of cognitive abilities that can be used with new participant samples, we prioritized increasing the reliability of the model over confirming theoretical relationships between cognitive abilities and game indicators.

To further reduce the overfitting of the model beyond what can be achieved by the regularization performed by the elastic-net model we used 100 times repeated 5-fold cross-validation (Burman, 1989). The standardization and imputation is performed separately for the training set in each of the cross-validations. The trained models ($\{\beta_{1j}$,...,$\beta_{45j}\}$, $k_j$) (see Supplemental Information Section 12) are the result of fitting the elastic net model to the entire training set with the best hyper parameters determined by the cross-validation. If a single game indicator or the cognitive ability measured by tasks was more than 3 SD's from the mean, the player was excluded from the fitting, as the fitting would be sensitive to such outliers.

We utilized the scikit learn library (Pedregosa et al., 2011) with Python 3.8.13 to perform the imputation, fit the elastic net model and perform the cross validation. Scikit learn defines the hyper-parameters of the elastic net model that control the regularization ($\alpha$, $L_1$) such that $\lambda_1 = \alpha L_1$and $\lambda_2 = \alpha(1 - L_1)$. For the elastic net hyper-parameter tuning we used the recommended $L_1$ ratios (0.1, 0.5, 0.7, 0.9, 0.95, 0.99, and 1), and let the elastic net function determine the appropriate $\alpha$ (*Sklearn.Linear_model.ElasticNetCV*, 2022).

### 3.4 Factor analysis of cognitive abilities from games and validation tasks

To identify the extent to which our predictive models rely on a generalized cognitive ability we perform an exploratory factor analysis on the cognitive abilities measured from the tasks and predicted from the games.

For this we apply the FactorAnalyser library (Biggs & Madnani, 2022) using principal factor extraction without any rotation. To evaluate the similarity of the main factor loadings ($F_{g,i}$ and $F_{t,i}$) we compute the cosine similarity $\cos(\theta) = \frac{\sum F_{g,i} F_{t,i}}{|F_g||F_t|}$. If the cosine similarity is 0 the main factors are orthogonal, and if it is 1 they are completely identical.

## 4. Results

### 4.1 Reliability of task measured cognitive abilities

The cognitive abilities measured from the task are formed by averaging the theoretically chosen reflective indicators with equal weights. Thus, the internal reliability of the cognitive abilities can be assessed by computing the Cronbach α. The Cronbach α of all but Visual Processing is above 0.7 (Table 2), which indicates good reliability. Categorical Visual Perception and Color Perception both contain only one indicator. Thus, a Cronbach α cannot be computed.

### 4.2 Cognitive modelling

The fitting and cross-validation process resulted in 8 accepted ($r_{cv} > 0.2$) prediction models with medium to strong effect sizes and five rejected models (Table 3). This cut-off turned out to align with whether the model significantly predicts more than an intercept-only model ($p_{cv} < 0.05$). More specifically, we accepted models of choice reaction time, categorical visual perception, central executive functioning, simple reaction time, response inhibition, visual processing, cognitive flexibility, and visual working memory. The coefficients of the models and brief interpretations of the relationships between game indicators and cognitive abilities can be found in Supplementary Information Section 12.

*Table 2 Cronbach α for the task measured cognitive abilities*

| Cognitive Ability | n | Cronbach α |
|---|---|---|
| Central Executive Functioning | 383 | 0.73 |
| Written Language Comprehension | 426 | 0.96 |
| Visual Processing | 313 | 0.25 |
| Visual Working Memory | 276 | 0.80 |
| Simple Reaction Time | 233 | 0.75 |
| Choice Reaction Time | 90 | 0.83 |
| Response Inhibition | 147 | 0.70 |
| Cognitive Flexibility | 222 | 0.70 |
| Planning | 198 | 0.71 |
| Color Perception | 426 | N/A |
| Categorical Visual Perception | 1199 | N/A |
| Recognition | 237 | 0.70 |
| Mental Rotation | 446 | 0.90 |

The cutoff at 0.2 for the estimated out-of-sample prediction strengths might seem like a low bar; however, the estimates are conservative compared to the full sample correlation. To remove the bias from overfitting the data in the full models' correlation with the tasks we estimated an *out-of-sample prediction strength* ($r_{cv}$, Table 3), i.e., what the correlation between the model predicted and the task measured cognitive abilities would be in an entirely new dataset. The estimate is the average correlation between the model predictions and the task-measured cognitive abilities on the test samples for each of the repeated cross-validation test sets. If we were to evaluate the models in a less conservative manner, all but one of the full sample correlations between the game predicted and task measured cognitive abilities (r, Table 3) would be medium to very-strong correlations (Cohen, 1988).

*Table 3 Results of fitting the cognitive abilities with an elastic-net model. The column r represents full sample correlations, whereas $r_{cv}$ (the grey column) is the estimated out-of-sample prediction strength from the repeated cross-validation. A negative value of $r_{cv}$ means that the model has no predictive power. $p_{cv}$ is the test of the cross validated out-of-sample prediction strength against an intercept-only model, i.e., is at least one coefficient significantly non-zero, and MAE is the mean absolute error. All cognitive abilities where standardized (M = 0, SD = 1). We accepted models for 8 of the 13 cognitive abilities (bold text).*

| Cognitive Ability | n | r | $r_{cv}$ | 95% Confidence Interval for $r_{cv}$ | $p_{cv}$ | MAE |
|---|---|---|---|---|---|---|
| Choice Reaction Time | 60 | 0.80 | 0.60 | [0.41, 0.74] | <.001 | 0.389 |
| Central Executive Functioning | 278 | 0.62 | 0.55 | [0.46, 0.62] | <.001 | 0.595 |
| Simple Reaction Time | 160 | 0.69 | 0.54 | [0.42, 0.64] | <.001 | 0.541 |
| Categorical Visual Perception | 855 | 0.54 | 0.51 | [0.46, 0.56] | <.001 | 0.605 |
| Response Inhibition | 99 | 0.61 | 0.45 | [0.28, 0.59] | <.001 | 0.518 |
| Visual Working Memory | 220 | 0.48 | 0.39 | [0.28, 0.50] | <.001 | 0.586 |
| Cognitive Flexibility | 152 | 0.51 | 0.28 | [0.13, 0.42] | <.001 | 0.599 |
| Visual Processing | 195 | 0.39 | 0.21 | [0.08, 0.34] | .003 | 0.710 |
| Color Perception | 296 | 0.19 | 0.11 | [-0.01, 0.22] | .063 | 0.698 |
| Written Language Comprehension | 296 | 0.21 | 0.06 | [-0.05, 0.18] | .265 | 0.689 |
| Mental Rotation | 318 | 0.35 | 0.03 | [-0.08, 0.14] | .630 | 0.599 |
| Planning | 142 | 0.30 | -0.05 | [-0.22, 0.11] | .521 | 0.544 |
| Recognition | 163 | N/A | -0.06 | [-0.21, 0.10] | .475 | 0.314 |

#### *4.2.1 Model generalizability with data from computer version*

The analysis above only considered the players on the mobile devices as the interface differences to computers could affect the measurement of the cognitive abilities (Drucker et al., 2013; Muender et al., 2019; Watson et al., 2013). However, if the accepted models that we have trained on the data from the mobile devices represents a mapping between cognitive abilities and game indicators, then applying them on the data collected from players of the computer version will provide a test of the generalizability. To account for systematic interface differences, we standardized the game and task indicators anew with only the computer data. We then computed the cognitive abilities from the task indicators for all the players that had taken the right combinations and correlated them with the predicted cognitive abilities from the models trained on mobile data (Table 4).

The eight accepted mobile-trained models predict the cognitive abilities for the players of the computer version, with correlation strengths of similar and on average higher magnitude as the out-of-sample prediction strength. This test constitutes powerful support for the reliability of the models.

*Table 4 Correlation between predicted and measured cognitive abilities for players on the computer version using models trained on data from the mobile version.*

| Cognitive Ability | n | r | $r\text{-}r_{cv}$ | 95% Confidence Interval for r | p |
|---|---|---|---|---|---|
| Choice Reaction Time | 49 | 0.68 | 0.09 | [0.50, 0.81] | <.001 |
| Central Executive Functioning | 137 | 0.52 | -0.02 | [0.39, 0.64] | <.001 |
| Simple Reaction Time | 97 | 0.60 | 0.06 | [0.46, 0.71] | <.001 |
| Categorical Visual Perception | 516 | 0.44 | -0.07 | [0.37, 0.51] | <.001 |
| Response Inhibition | 72 | 0.64 | 0.19 | [0.48, 0.76] | <.001 |
| Visual Working Memory | 141 | 0.36 | -0.03 | [0.21, 0.50] | <.001 |
| Cognitive Flexibility | 93 | 0.37 | 0.08 | [0.18, 0.53] | <.001 |
| Visual Processing | 102 | 0.40 | 0.18 | [0.22, 0.55] | <.001 |

## 4.3 Assessing the models' predictive power

### *4.3.1 Generalized cognitive ability*

Since the cognitive abilities are related in the theoretical framework (Fig 1), it is essential to look at shared variation contributing to the observed predictive power. Therefore, we performed a pair of exploratory factor analyses, one on the cognitive abilities computed from validation tasks, and one on cognitive abilities predicted from game data. This allowed us to identify the main factor in both sets, interpretable as a generalized cognitive ability (Knopik et al., 2017).

The factor analysis's exclusion criterion was whether the cognitive ability measure was more than 3 SD's from the population mean. This criterion was different from the one applied during the fitting procedure, as a single outlier among the game indicators could potentially be compensated for in the predictive model, either by all the other non-outliers or that a particular game indicator is irrelevant for that particular model. Thus, we decided to exclude based on the predicted value rather than at the game indicator level. The same criterion is used for all the following analyses in this paper. This meant that, for cognitive abilities measured by games, the factor analysis included 6,546 players. For cognitive abilities measured by validation tasks, 82 out of the 84 players with all cognitive abilities measured by the tasks were included. The relatively low task participant number reflects that the completion of all 14 validation tasks were required to be included in the analysis.

Results of the factor analyses revealed that, for both game-based and validation-based measures, the components in the framework are not orthogonal, and unsurprisingly, there is a large shared main factor across all cognitive abilities (Fig 6)

The fact that the percentage of variance explained is higher for games (Fig 6a) was expected, since the number of indicators used to evaluate the cognitive abilities had decreased from 68 task indicators to 45 game indicators. Therefore, if we make a simplified evaluation of the amount of absolute rather than the proportional variance explained by the main factors it is 36.72 (68 task indicators with unit variance · 54%) for the tasks and 37.35 (45 task indicators with unit variance · 83%) for the games, which is similar to each other.

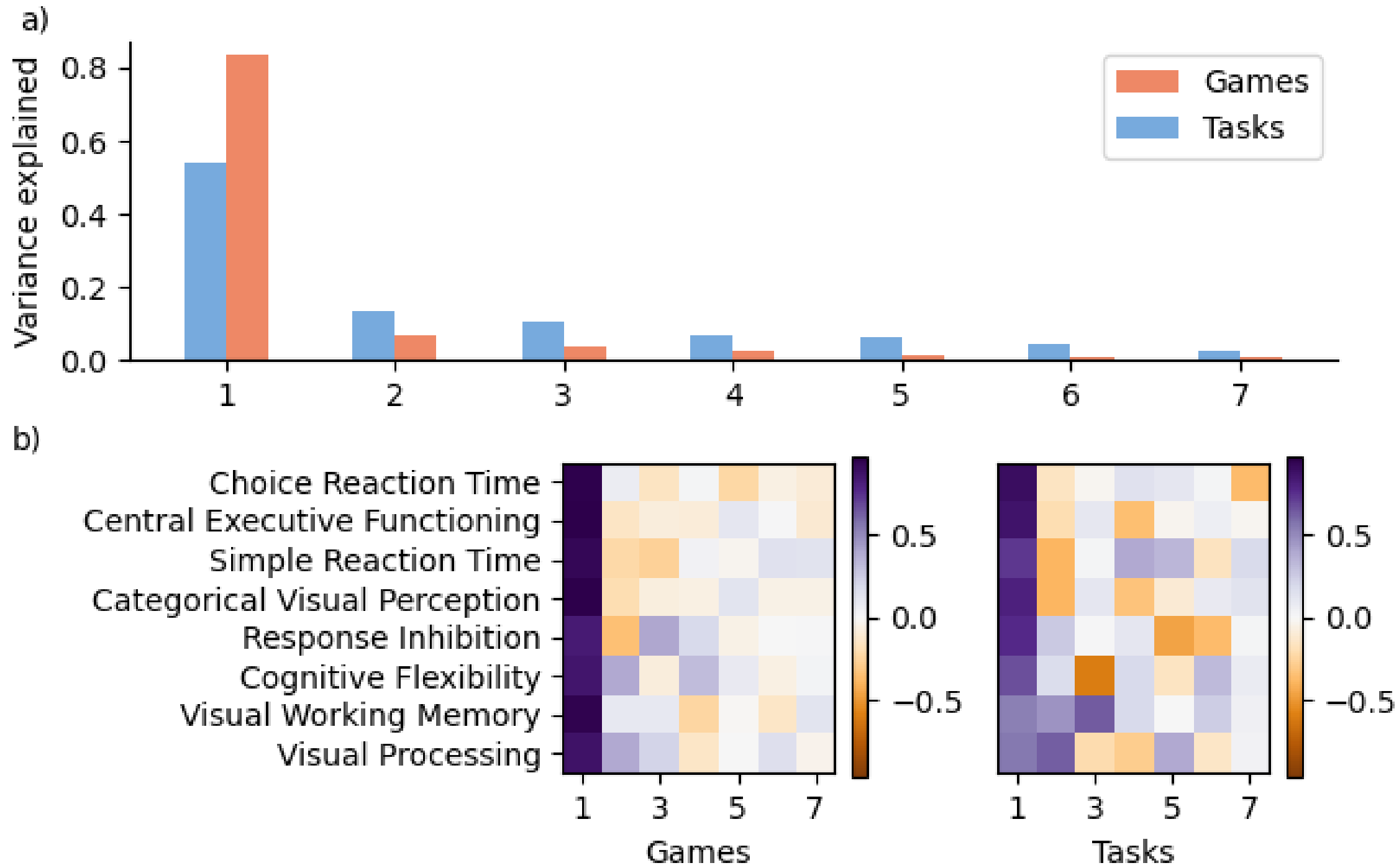

*Fig 7 a) Proportion of variance covered by each factor. b) Loadings of each cognitive ability on the factors.*

The main factor loadings are very similar across all cognitive abilities, with our predictive game-based model yielding similar results as the validation tasks (cosine similarity = 0.98). For both games and tasks, the main factor corresponds approximately to the mean of all the cognitive abilities (Fig 6b).

#### *4.3.2 Discriminant validity of the models*

As shown above, the main factor is responsible for explaining a high percentage of variance for both games and validation tasks. Therefore, in order to demonstrate that our model has discriminative power beyond being driven by the main factor, we computed partial correlations between the games and validation tasks while controlling for the games' main factor. These partial correlations thus reveal the extent to which our models can predict the nuances contained within each separate cognitive ability that goes beyond a generalized cognitive ability. Fig 7 illustrates the fraction of the correlation between the task and the game-based measures that is not explained by the main factor. For all eight cognitive models, we find that 23-63% of the correlation is not due to the main factor, demonstrating the discriminative validity of the models. In other words, we clearly document that each of our models tap significantly into aspects beyond just the general abilities factor.

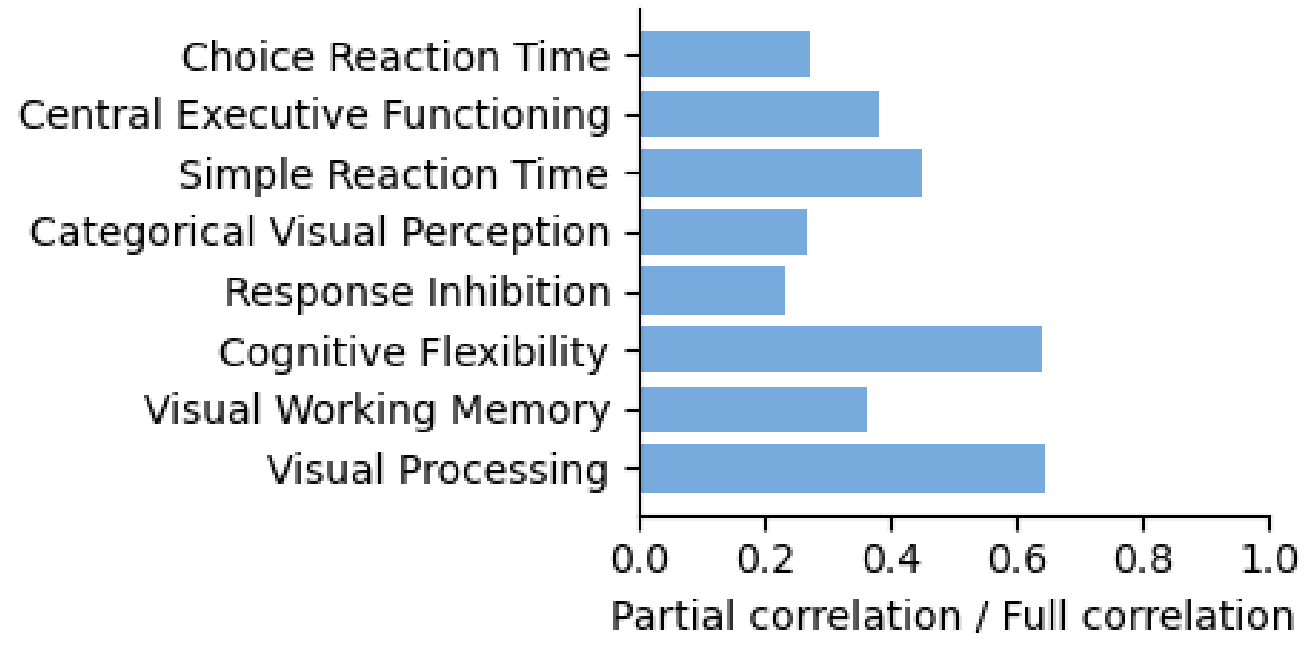

*Fig 6 The proportion of the models' predictive strength not explained by the main factor. The full correlations are similar to, but not exactly equal to, the r values found in the Table 3. A table with values of the full and partial correlation can be found in the Supplementary Information Section 16.*

## 4.4 Skill Lab as a potential cognitive diagnostics tool

One of Skill Lab's potential use cases is as a low-cost test battery that could be used to track cognitive impairments. We are therefore interested in the time it takes compared to current cognitive batteries. The average time taken to complete all the six games was 14 minutes (SD = 5 min), in comparison with 72 minutes needed to complete all the validation tasks (SD = 7). In other words, the Skill Lab games could model cognitive abilities in one-fifth of the time as required by the traditional set of cognitive tests.

To further demonstrate the potential of Skill Lab as a diagnostic tool, we use the trained models to illustrate the cross-sectional cohort distributions of cognitive abilities by age for the Danish population (Fig 8 and Fig SI. 45-53). Examining the distributions obtained from the games across ages, we observed the expected increase in all cognitive abilities from age 16 to 20 years, followed by a gradual decline from age 20 years.

## 5. Discussion

We designed the Skill Lab games to simultaneously engage and measure multiple cognitive abilities in a more realistic gameplay context within a single convenient, engaging, and scalable package. One of this project's main contributions is a demonstration that we were able to achieve a large-scale *in-the-wild* within-sample validation of our cognitive assessment. We first constructed predictive models of cognitive abilities based on data from 1,351 participants who had completed a sufficient number of both games and tasks, then validated the performance of these models based on data from 6,369 players who played at least one game on mobile devices. We were also able to further validate our model based on the data from the in total 603 players who played the game and took the validation tasks on computers.

It should be mentioned that there was no nudging towards the tasks within Skill Lab and no requirement to do so; thus, there were no expectations from a data collection perspective towards the in-the-wild players completing all tasks. In particular, it should be noted that such a large fraction of the players identified sufficiently with the scientific purpose of the games (to help the researchers better understand human cognition) that they spent so much time performing the rather tedious validation tasks without any form of extrinsic reward. Our study achieves both exemplary breadth of different abilities and depth of volunteer participation compared to other game-based population-scale assessment studies such as SeaHero Quest and The Great Brain Experiment (H. R. Brown et al., 2014; Coughlan et al., 2019; Coutrot et al., 2018; Hunt et al., 2016; McNab et al., 2015; McNab & Dolan, 2014; Rutledge et al., 2014, 2016; Smittenaar et al., 2015; Teki et al., 2016). This is a positive step towards comprehensive citizen involvement in the construction of complex cognitive studies in the future.

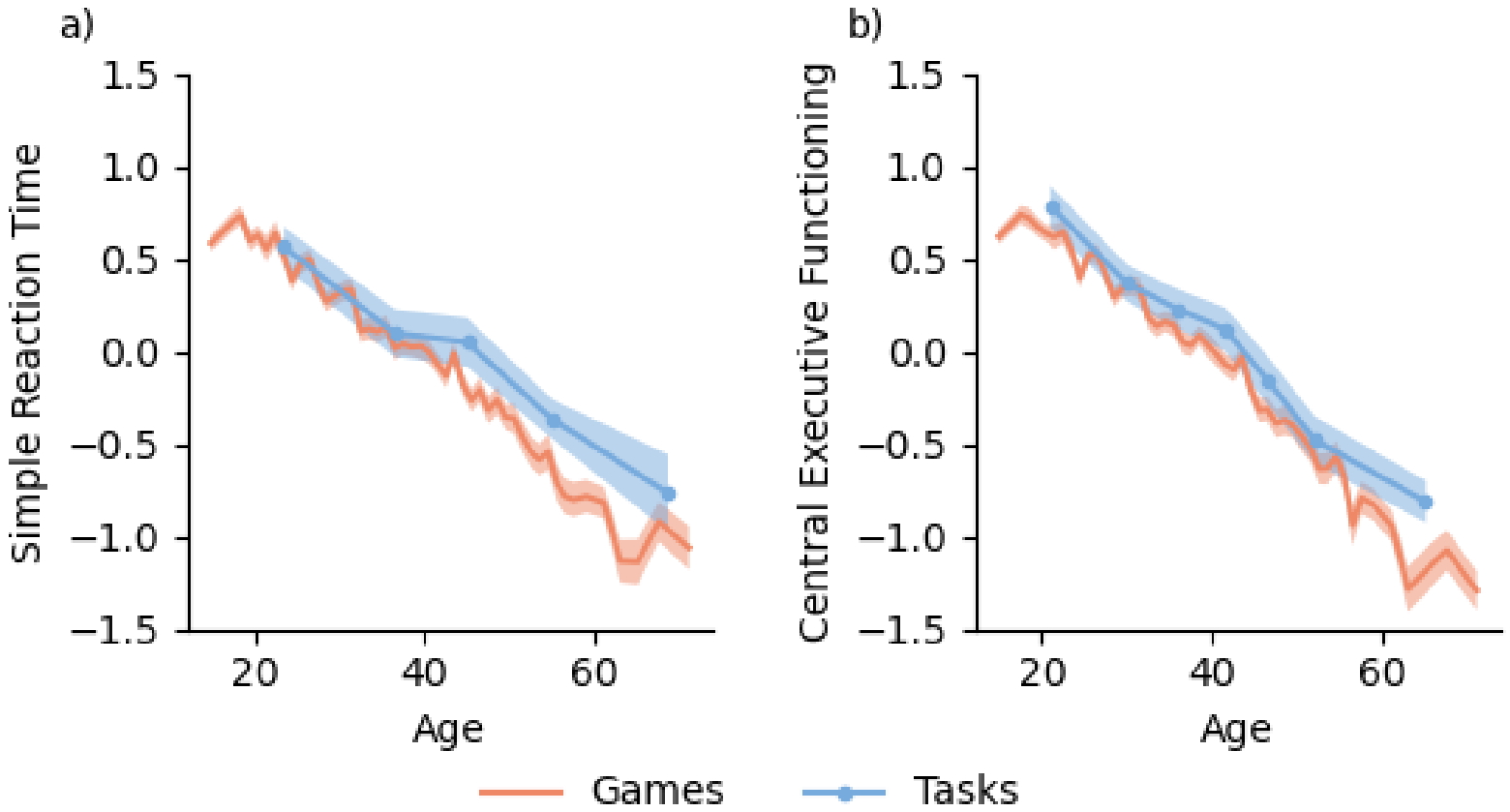

*Fig 8 Cognitive abilities across age groups for a) Simple reaction time($n_{task}$ = 225, $n_{wild}$ = 6277) and b) Central executive functioning ($n_{task}$ = 372, $n_{wild}$ = 6281). The shaded areas around the curves are the standard error of the mean. The y-axis represents simple reaction time and central executive functioning standardized across the population, thus, higher values on the y-axis corresponds to faster reaction times (the curves for the remaining cognitive abilities can be found in the Supplemental Information Chapter 14). Each age point in the graph includes at least 30 players. The points were generated by starting at age 16 and checking whether 30 players of that age whose data provided a cognitive ability measure. If there were enough players, the following point was generated starting with those one year older; if not, the following ages were added one year at a time until a sample size of 30 was reached.*

In line with the goals of our design process, results from the study demonstrated good convergent validity of the game-based cognitive measures, where eight of the models predicting the cognitive abilities from game indicators correlated well with the task-based measures. The factor analysis revealed a main factor for cognitive abilities that could be interpreted as a general cognitive ability for both games and tasks (Fig 6) in line with a priori expectations during the design phase (Fig 1). Via partial correlations (Fig 7), we demonstrated that the shared information from the main factor is insufficient to explain a substantial proportion of each cognitive ability's observed agreement between task and game estimates. Each of our measures, therefore, captures some of the nuances of the cognitive abilities beyond the dominant factor.

### 5.1 Limitations

While showing exciting potential for future applications, our current study is limited in that people were only recruited to play the game once. In order to be considered as a potential clinical tool in one-off as well as longitudinal applications, a follow-up test-retest study is needed to assess the robustness of our cognitive ability estimates. In such a test-retest setup, we could control the time between playthroughs to neutralize learning effects and ensure all the games have been played in both playthroughs. It is not unreasonable to expect that we could achieve even more consistent estimates by training models dependent on the playthrough number, compensating for learning effects due to the player familiarizing themselves with the tasks. Another consideration regarding reliability is the fact that our validation population set exhibited a slightly different gender distribution compared to the overall dataset (64% versus 49% females overall, see Supplementary Information Section 9 for full demographic breakdown).

In addition, our sample population, while diverse in age, comes primarily from Denmark. If we want to establish more general demographic norms than those we have collected on the Danish population, we would naturally have to expand our recruitment efforts. As part of these efforts, we have prepared a Spanish translation of Skill Lab in addition to the Danish and English translations that already existed, with the plans to launch the game internationally in the future.

### 5.2 Future directions and applications

As an example of what our Skill Lab models are currently able to do, we used our population sample to replicate previous findings regarding the age distribution of cognitive abilities. Our study offers a cross-sectional snapshot of the Danish population, comprising the largest open normative dataset of these cognitive abilities. The observed patterns (Fig 8) follow the previously established expectations (Lindenberger, 2014; Salthouse, 2019), which supports Skill Lab's validity as an assessment tool. This of course comes with the caveat that the age-related decline we observed in the present study could have also been confounded by factors such as technological familiarity, which we did not measure. With appropriate future work to account for such confounds, our dataset may serve as a normative benchmark for future applications, not only within psychology but also for the social sciences, clinical applications, and education. These finely stratified age norms will be of particular importance when Skill Lab addresses questions that require age-based controls.

An alternative to the computational approach we present in this paper of aggregating indicators from multiple tasks is testing the feasibility of predicting individual task indicators from game data, which is more in line with the conventional literature (Salthouse, 2011). However, predicting individual indicators is not very robust, so we made the pragmatic choice of defining aggregated cognitive abilities measures (Bollen & Bauldry, 2011) while only combining task indicators associated with a cognitive ability in the theory to strengthen its interpretation. The eight accepted models already represent a broad, strong, and rapid testing battery. We exposed these choices to potential disconfirmation in the current work by examining their agreement across independent estimates; rejecting three of thirteen while accepting ten. Since the data set is open, it is also open for potential explorations of alternative choices. We have taken preliminary steps in this direction by pursuing a theory-driven approach, in which we only include the game indicators that are theoretically associated with a specific cognitive ability during the fitting process. The results are qualitatively similar to the ones presented here but somewhat lower in quantitative effects as expected from a restricted model. Further work in this direction may help the iterative development toward games that are optimally suited for high-quality assessment of each ability.

In conclusion, the models developed through our work with Skill Lab illustrate the viability of a crowdsourcing approach in validating a cognitive assessment tool, which has several key implications. First, it allows scientists to create better human cognition models and test and validate cognitive abilities, potentially providing efficient ways to scale insights into particular cognitive abilities and how they are related to solving problems (Woolley et al., 2010). Second, we have generated a unique and open dataset, which includes normative benchmarks, that can be used as a basis for other studies. Finally, Skill Lab allows normative data for diverse populations, cultures, and languages to be collected in the future, facilitating the much-needed broadening of the samples typically tested in psychological and social science studies (Henrich et al., 2010). An advantage of Skill Lab over the traditional tests is that it is faster to play all six games once than to go through all the traditional cognitive tasks. Thus, the games could provide a low-cost self-administered test suitable for extensive deployment. This could be of great value to, e.g., the psychiatric sector in which current cognitive test batteries are burdensome to administer (Baune et al., 2018), leading to cognitive impairments often going unrecognized (Groves et al., 2018; Jaeger et al., 2006).

## 6. Acknowledgments

The authors acknowledge funding from the ERC, H2020 grant 639560 (MECTRL) and the Templeton, Synakos, Novo Nordic and Carlsberg Foundations. We would like to thank the Danish Broadcast Company DR for their collaboration without which the recruitment to the study would not have been as successful as it was. We would also like to thank the ScienceAtHome team and developers for making their contribution in designing and developing Skill Lab: Science Detective. Furthermore, we would like to acknowledge Susannah Goss for her help with copy editing, Michael Bang Petersen for commenting on the results, and Steven Langsford for his comments and help in the editing of the manuscript.

## 7. Availability of data, code, and materials

Skill Lab is available on the Apple App Store, Google Play, and online at (https://webgl.scienceathome.org/slsd/, Retrieved: 2023-06-06).

The raw and processed data that support the findings of this study are available together with the data processing scripts on the Open Science Framework (https://doi.org/10.17605/OSF.IO/PNW5Z , Retrieved: 2023-06-06).